\documentclass{article}[]
\usepackage{spconf}
\usepackage{graphicx,color}
\usepackage{amsmath,amssymb}
\usepackage{bm}
\usepackage{ascmac}
\usepackage{url}                      
\usepackage{hyperref}
\usepackage{ifthen}
\newcommand{\sign}[1]{{\sf sign}(#1)}
\newcommand{\tr}[1]{{\sf tr}(#1)}
\newcommand{\mmse}[1]{\eta_{m}(#1)}

\newcommand{\im }{\mathrm{j}}
\newcommand{\lH }{{H}}

\usepackage{comment}

\title{Complex trainable ISTA for linear and nonlinear inverse problems}

\name{Satoshi Takabe$^{\star \dagger}$ \qquad Tadashi Wadayama$^{\star}$ \qquad Yonina C. Eldar$^{\ddagger}$\thanks{This work was partly supported by
JSPS Grant-in-Aid for Scientific Research (A) Grant Number 17H01280,
JSPS Grant-in-Aid for Scientific Research (B) 
Grant Number 19H02138 (TW) and Grant-in-Aid 
for Early-Career Scientists Grant Number 19K14613, 
and the Telecommunications Advancement Foundation (ST).
e-mail: s\_takabe@nitech.ac.jp}
}
\address{$^{\star}$Department of Computer Science, Nagoya Institute of Technology, 
Nagoya, Aichi, Japan \\
$^{\dagger}$ RIKEN Center for Advanced Intelligence Project, Chuo-ku, Tokyo, Japan\\
$^{\ddagger}${Department of Math and Computer Science, Weizmann Institute of Science, Rehovot, Israel}			   
}

\begin{document}
%\ninept % for 9pt
%
\maketitle

\begin{abstract}%100-150 words
Complex-field signal recovery problems from noisy linear/nonlinear measurements appear in many areas of signal processing and wireless communications. 
In this paper, we propose a trainable iterative signal recovery algorithm named complex-field TISTA (C-TISTA) which treats complex-field nonlinear inverse problems. 
C-TISTA is based on the concept of deep unfolding and consists of a gradient descent step with the Wirtinger derivatives followed by a shrinkage step with a trainable complex-valued shrinkage function. Importantly, it contains a small number of trainable parameters so that its training process can be executed efficiently.
Numerical results indicate that C-TISTA shows remarkable signal recovery performance compared with existing algorithms. 
\end{abstract}


\section{Introduction}\label{sec_1}

{Inverse problems are prevalent} topics in signal processing and wireless communications.
For example, linear inverse problems {consider estimating}  inputs $\bm{x}\in\mathbb{R}^{n}$ from noisy  observations $\bm{y}=\bm{A}\bm{x}+\bm{w}\in\mathbb{R}^{m}$ {and can model, for example,} signal detection in multiple access systems with a channel matrix $\bm A$ such as multiple-input multiple-output (MIMO) and orthogonal frequency-division multiplexing (OFDM) systems~\cite{Tse}.
In these cases, we have prior information that the inputs take discrete values defined by a signal constellation. 
Compressed sensing (CS)~\cite{CS1,Eldar1, Eldar2} is another {popular} class of inverse problems widely applied to wireless communication techniques~\cite{Hayashi}.
CS enables {recovery of} sparse signals accurately even in underdetermined cases, i.e., $m<n$,
which is {the case, e.g.} in overloaded non-orthogonal multiple-access systems. 

Recently machine learning {methods have been shown to be} a powerful tool for inverse problems in wireless communications~\cite{ML}.
In particular, \emph{deep unfolding} {which originated} in learned {iterative soft thresholding algorithm} (LISTA)~\cite{LISTA} is a promising technique to {derive tailored deep networks} due to its high scalability and {ability to use prior information efficiently}~\cite{DU}.
Deep unfolding is {based on starting with iterative algorithms and then unrolling their recursive structures to deep networks} with trainable parameters. 
These parameters are trained by supervised signals using deep learning techniques.
A number of deep unfolding-based algorithms have been proposed:
{learned approximate message passing (LAMP)~\cite{LAMP} and}
\textit{trainable ISTA (TISTA)} for CS~\cite{TISTA2}, MIMO detectors~\cite{DMD1,TPG2}, OFDM signal detectors~\cite{JZ}, {trainable robust principle component analysis~\cite{RPCA}},  and decoders for error-correcting codes~\cite{Nach,LDPC}.
They are {mainly} based on signal recovery algorithms for \textit{real-valued linear} systems. 

A drawback of these algorithms is a limitation to
\emph{complex-field nonlinear inverse problems} with nonlinear observations $\bm{y}=f(\bm{A}\bm{x})+\bm{w}$, where $f:\mathbb{C}\rightarrow\mathbb{C}$ is an element-wise function. 
For complex-valued problems, a conventional transformation from complex vectors to real vectors is {typically
} used.
However, it possibly breaks correlations between real and imaginary parts of signals, which {can} degrade signal recovery performance. 
Adaptability to nonlinearity is another crucial issue. %in those algorithms.
Amplitude clipping and quantization are important in nonlinear CS~\cite{Yang}.
In wireless communications, amplitude clipping arises in OFDM systems {to reduce peak power~\cite{Han},} and quantization corresponds to the use of analog-to-digital converters. 

In this paper, we propose a trainable signal recovery algorithm to treat {complex-field nonlinear inverse problems} in a direct and general manner by deep unfolding.
The algorithm named complex-field TISTA (C-TISTA) is constructed as an extension of TISTA using 
the \emph{Wirtinger derivative} and complex-valued shrinkage function with trainable parameters. 
Numerical simulations demonstrate that C-TISTA outperforms existing algorithms in complex-valued CS and clipped OFDM signal detection.

{Throughout the paper, we use the following notation: 
For a function $f: \mathbb{C} \rightarrow \mathbb{C}$ and $\bm{v}=(v_1, \dots, v_n)^T  \in \mathbb{C}^n$, 
we define $f(\bm{v}) := (f(v_1), \dots, f(v_n))^T$.
For a vector $\bm{z}$,
$\bm{z}^\ast$ represents its conjugate. %vector.
For a matrix $\bm{A}:=(A_{i,j})\in\mathbb{C}^{m\times n}$, $\bm{A}^\lH := (A_{j,i}^\ast)$ is 
its Hermitian transpose.
We define $\mathcal{CN}(\mu, \sigma^2)$ as a complex Gaussian distribution with mean $\mu$ and variance $\sigma^2$.
The p.d.f. of $\mathcal{CN}(\mu, \sigma^2)$ is defined by
\begin{equation}
G(x;\mu, \sigma^2) := \frac{1}{\pi \sigma^2}\exp\left(-\frac{|x-\mu|^2}{\sigma^2}\right). \label{eq_GLM}
\end{equation}
For a random variable $\bm{x}\in\mathbb{C}^n$ and a function $g:\mathbb{C}^n\rightarrow\mathbb{C}^n$, 
$\mathsf{E}_{\bm x}g(\bm{x})$ denotes the expectation of $g(\bm{x})$ with respect to $\bm{x}$.
}

\section{Brief review of TISTA}\label{sec_2}

We first briefly review TISTA for 
real-valued linear observations defined by 
\begin{equation}
\bm{y} = \bm{A} \bm{x} + \bm{w},\label{eq_lin}
\end{equation}
where $\bm{A} \in \mathbb{R}^{m \times n} (n \ge m)$ and $\bm{x}\in \mathbb{R}^n$. 
Each entry of the noise vector $\bm{w} \in \mathbb{R}^{m}$ 
follows a zero-mean Gaussian distribution with variance $\sigma^2$.
We also assume that the prior information on $\bm{x}$ is known.

{For a sparse input $\bm{x}$, LASSO formulation~\cite{LASSO} is conventionally used for its estimation, leading to 
\begin{equation}
\bm{\hat{x}} =\mathrm{minimize}_{\bm{x}\in\mathbb{R}^n} \|\bm{y} - \bm{A} \bm{x}\|_2^2 + \lambda \|\bm{x}\|_1,
\label{eq_LASSO}
\end{equation}
where $\|\cdot\|_1$ is the $\ell_1$ norm and $\lambda$ is a positive constant.
}

{ISTA~\cite{ISTA} is a simple iterative algorithm to solve (\ref{eq_LASSO}).
It consists of the following recursions:
\begin{equation} \label{ISTA_gradient}
\bm{s}^{(t + 1)} := \eta_s(\bm{s}^{(t)} + \beta \bm{A}^T(\bm{y} - \bm{A} \bm{s}^{(t)}); \tau),
\end{equation}
where $\beta (>0)$ is a step size 
and  
$\eta_s (r; \tau)  = \sign{r} \max \{ |r| - \tau, 0 \}$ is the element-wise soft thresholding function with threshold 
$\tau \in \mathbb R$ $(\tau > 0)$ related to $\lambda$.
%The update equation (\ref{ISTA_gradient}) is said to be the gradient descent step and
%(\ref{ISTA_shrinkage}) is said to be the proximal step.
Since the soft thresholding function is the proximal operator of the $\ell_1$-regularizer, 
ISTA can be seen as a proximal gradient descent algorithm for solving (\ref{eq_LASSO}). 
Note that, in order to have convergence, 
the step size $\beta$ should be carefully chosen~\cite{ISTA}.
}

{LISTA~\cite{LISTA} is a trainable algorithm based on ISTA.
As deep unfolding, the signal-flow graph of ISTA is expanded to a deep network
 with embedded trainable parameters such as a step size. 
The recursive formula is given by 
\begin{equation} 
\bm{s}^{(t + 1)} := \eta_s(\bm B_t \bm{y} + \bm C_t \bm{s}^{(t)}; \tau_t), \label{eq_LISTA}
\end{equation}
where $\{\bm B_t, \bm C_t,\tau_t\}_{t=1}^T$ is a set of trainable parameters.
%Note that the special case in which $\bm B_t= \beta \bm  A^T$, $\bm C_t= \bm I_n -\beta \bm  A^T\bm  A$,
%and $\tau_t=\tau$ corresponds to the original ISTA.
These parameters are learned by back propagation and stochastic gradient descent using training data.
Although {the convergence speed of LISTA is significantly faster than
 that of ISTA, its training process is costly and 
sometimes unstable because the number of trainable parameters is {large, i.e.,} $(mn+n^2+1)T$ in $T$ iterations.
}

TISTA~\cite{TISTA2} is another trainable algorithm for the system (\ref{eq_lin}), 
which is based on orthogonal AMP~\cite{OAMP} and  defined by
\begin{align}
\bm{r}^{(t)} &:= \bm{s}^{(t)} + \gamma_t \bm{W} (\bm{y} - \bm{A} \bm{s}^{(t)}), \label{lmse}\\
\bm{s}^{(t+1)} &:= \mmse {\bm{r}^{(t)}; \tau_t^2},   \label{mmse}\\
v_t^2 &:= \max \left\{ \frac{\|\bm{y} - \bm{A} \bm{s}^{(t)}\|_2^2 - m \sigma^2}{\tr{\bm{A}^T \bm{A}}}, \epsilon \right\}, \label{v_t}\\ 
\tau_t^2 &:= \frac{v_t^2}{n} (n + (\gamma^2_t  - 2\gamma_t)  m   )
+ \frac{\gamma_t^2\sigma^2}{n} \tr{\bm{W} \bm{W}^T}, \label{tau_est}
\end{align}
where the matrix $\bm{W} \!=\! \bm{A}^T(\bm{A} \bm{A}^T)^{-1}$
is the pseudo-inverse matrix 
of $\bm{A}$.
The shrinkage function $\mmse{\cdot}$ is an {element-wise minimum mean square error (MMSE) shrinkage function whose shape is similar to the soft shrinkage function.}
The parameter $\gamma_t$ is a trainable parameter that can be 
optimized in training process.
{TISTA has only $T$ trainable parameters in $T$ iterations.
This fact} leads to scalable and stable training process {compared with LISTA.} 
{In addition, TISTA can improve signal recovery performance and adaptability to wide range of measurement matrices in CS~\cite{TISTA2}.}

{In this paper, we propose a TISTA-based algorithm for complex-field nonlinear systems. 
We will modify (\ref{lmse})-(\ref{tau_est}) using the Wirtinger derivative and a trainable shrinkage function
corresponding to various prior information including sparsity and discreteness.}

\section{Complex-field TISTA}\label{sec_4}

\subsection{System Model}
%In the following sections, 
We consider a complex-field nonlinear system given by 
\begin{equation}
\bm{y} = f(\bm{A}\bm{x}) + \bm{w}, \label{eq_GLM}
\end{equation}
where $\bm{A} \in \mathbb{C}^{m \times n}$ is a measurement matrix.
The vector $\bm{x} \in \mathbb{C}^n$ is %called 
the input vector on which a certain 
prior information such as sparsity or discreteness is imposed. 
Each component of the the noise vector $\bm{w} \in \mathbb{C}^m$
follows ${\cal CN}(0, \sigma^2)$. %with variance $\sigma^2$.
The inverse problem is the estimation the input from 
the observation $\bm{y}$ based on
 $\bm{A}$ and $f$.

\subsection{Complex-field TISTA}

We now propose C-TISTA for complex-field 
nonlinear inverse problems. 
We apply {the} Wirtinger derivative %in the gradient step of TISTA
and trainable complex-valued shrinkage function to TISTA. 
The recursive formulas of C-TISTA are given as follows:
{\begin{align}
\bm{r}^{(t)} &:= \bm{s}^{(t)} + \beta_t h(\bm{s}^{(t)}), \label{eq_tista_lin}\\
\bm{s}^{(t+1)} &:= C_t (\eta(\bm{r}^{(t)}; \lambda^{(t)}) - D_t\bm{r}_t), \label{eq_tista_non}\\
h(\bm{s})& := \bm{V}
\left[ \{\bm{y} - f(\bm{A}\bm{s})\}^\ast \odot \frac{\partial f}{\partial z^\ast}(\bm{A}\bm{s})\right.\nonumber\\
&\left.\qquad\qquad\quad+\{\bm{y} - f(\bm{A}\bm{s})\} \odot \frac{\partial f^\ast}{\partial z^\ast}(\bm{A}\bm{s})\right],\label{eq_tista_grad} 
\end{align} 
where $\eta(z;\lambda): \mathbb{C}\rightarrow \mathbb{C}$ is a nonlinear function with a parameter $\lambda$
and $\odot$ is the Hadamard product.
\begin{comment}
{the Wirtinger derivative of a function $f:\mathbb{C} \rightarrow \mathbb{C}$ in (\ref{eq_tista_grad})  is defined by
\begin{align}
\frac{\partial f(z)}{\partial z} &:= \frac{1}{2} \left[ \frac{\partial}{\partial {z}_{\mathrm{r}}}-\im \frac{\partial}{\partial {z}_{\mathrm{i}}} \right] F({z}_{\mathrm{r}},{z}_{\mathrm{i}}), 
\\
\frac{\partial f(z)}{\partial z^\ast} &:= \frac{1}{2} \left[ \frac{\partial}{\partial {z}_{\mathrm{r}}}+\im \frac{\partial}{\partial {z}_{\mathrm{i}}} \right] F({z}_{\mathrm{r}},{z}_{\mathrm{i}}), 
\end{align}
where $F:\mathbb{R}^2 \rightarrow \mathbb{C}$ is defined by 
$F({z}_{\mathrm{r}},{z}_{\mathrm{i}}) = f(z)$ for ${z}_{\mathrm{r}}=\mathrm{Re}(z)$ and ${z}_{\mathrm{i}}=\mathrm{Im}(z)$.}
\end{comment}
{The definition of the Wirtinger derivative and derivation of (\ref{eq_tista_grad}) can be found in Appendix~\ref{sec_app}.}
As a matrix $\bm V$, we can use $\bm A^{\lH}$ or the pseudo-inverse matrix 
$\bm W:=(\bm{A}^\lH \bm{A})^{-1}\bm{A}^\lH $ ($m\le n$) similar to TISTA.
Starting from an initial point $\bm{s}^{(1)}$,
the estimate after $T$ iterations is given by $\bm{\hat{x}}:=\bm s^{(T+1)}$.
The trainable parameters of C-TISTA are $4T$ real scalars $\{\beta_t,\lambda_t,C_t,D_t\}_{t=1}^T$,
which is constant to $n$ and $m$ and  leads to scalable and stable training process.  

%In the numerical experiments in the following sections, we use $\bm{s}^{(1)} :=\bm{W} \bm{y}$}, which is the estimate
% by the zero-forcing (ZF) detector.
 
%\subsection{Details of C-TISTA update rules} 
{The first update rule (\ref{eq_tista_lin}) is a \emph{gradient step} using the Wirtinger derivative
whose step size is a trainable parameter $\beta_t(>0)$.
In the second update rule (\ref{eq_tista_non}) called a \emph{shrinkage step}, 
an estimate is updated by a trainable divergence free (DF)-like function~\cite{OAMP, He2} with shrinkage function $\eta(\cdot)$ to reflect prior information on $\bm{x}$ and trainable parameters $\{\lambda_t,C_t,D_t\}$.

In this paper, we use two shrinkage functions depending on prior information.
The first one is the complex-valued soft shrinkage function for sparse prior. It is defined by
\begin{equation}
	\eta_{{cs}}(x; \lambda) := \eta_s(x; \lambda) e^{\im \varphi(x)}, \label{eq_soft}
\end{equation}
where $\varphi(x)$ is the phase of $x$ and 
 $\lambda(>0)$ means the threshold~\cite{CAMP}.
The second one is an MMSE estimator for discrete signals.
Let $x$ be a random variable uniformly chosen from a signal constellation 
$S :=\{s_1, s_2, \ldots, s_M \} \subset \mathbb{C}$ of size $M$.
For a {virtual AWGN channel} $p(y|x) = G(y;x,\lambda)$ with variance $\lambda$,
the MMSE function is given by
\begin{equation}
	\eta(y; \lambda) := {\sf E}_{{x}}[x|y] = \frac{\sum_{s \in S} 
	s \exp\left( - \frac{|y - s|^2}{ \lambda}\right) }{\sum_{s \in S} 
	\exp\left( - \frac{|y - s|^2}{ \lambda}\right)}.\label{eq_sh}
\end{equation}

\section{Numerical Results}\label{sec_5}
{We here examine the recovery performance of C-TISTA in complex-valued CS and clipped OFDM signal detection}.

\subsection{Implementation Details}
C-TISTA is implemented by PyTorch 1.2~\cite{PyTorch}.
{We set an initial point to $\bm s^{(1)} \!=\!\bm W\bm y$ and 
use the pseudo-inverse matrix $\bm W$ as $\bm V$ in (\ref{eq_tista_grad}).}  
The training parameter $\lambda_t$ is replaced {by} $a_t^2$ to satisfy the identity $\lambda_t\!>\!0$. 
Training process is executed by incremental training similar to TISTA~\cite{TISTA2} to avoid a vanishing-gradient problem.
Adam optimizer~\cite{Adam} with learning rate $0.005$ is used to minimize {the} MSE loss function between the true signal $\bm{x}$ and estimate $\bm {\hat{x}}$.    
{The parameters are initialized as} $(\beta_t, a_t, C_t, D_t)=(0.1,1,1,0)$ for $t=1,\dots,T$.

\begin{figure}[t]
   \centering
   \includegraphics[width=0.82\hsize]{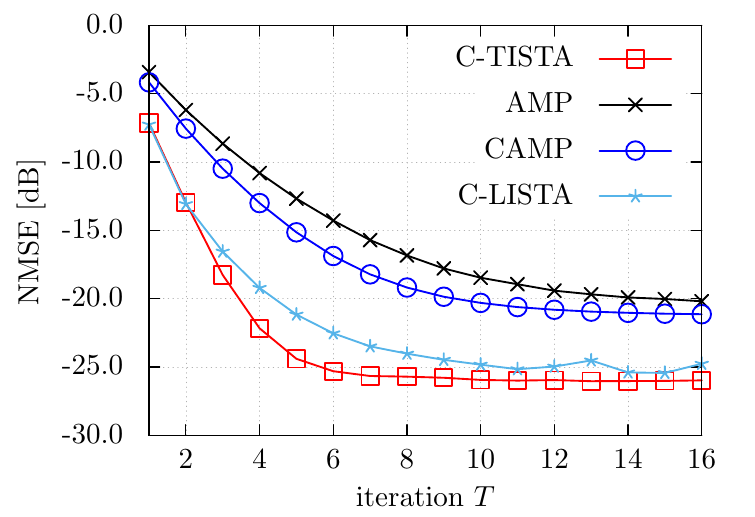}
    \caption{{Signal recovery in complex-valued CS ($n\!=\!240$, $m\!=\!120$, $p\!=\!0.1$).}
    NMSEs are plotted as functions of iteration $T$.  Symbols represent C-TISTA (squares), AMP (cross marks), CAMP (circles), and C-LISTA (asterisks).
    }
    \label{fig_cs}
\end{figure}

\begin{figure*}[t!]

\begin{minipage}[b]{.49\linewidth}
  \centering
  {\includegraphics[width=0.78\textwidth]{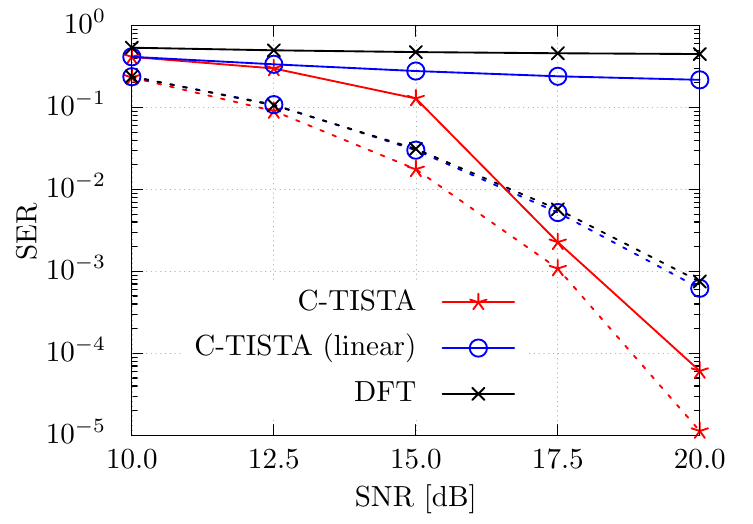}}\\
%  \vspace{1.5cm}
  {(a) SER as a function of SNR. }\medskip
\end{minipage}
\hfill
\begin{minipage}[b]{.49\linewidth}
  \centering
  {\includegraphics[width=0.78\textwidth]{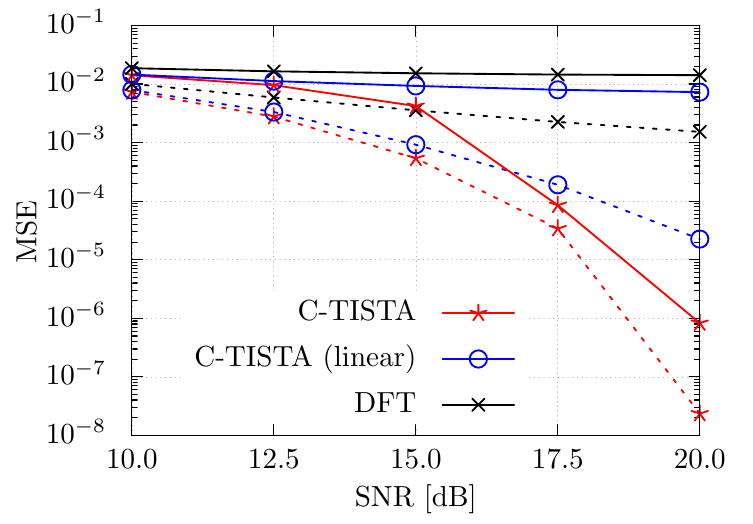}}\\
%  \vspace{1.5cm}
  {(b) MSE as a function of SNR. }\medskip
\end{minipage}
\caption{Signal detection in clipped OFDM systems ($n=192$). Symbols represent C-TISTA (stars), C-TISTA without nonlinearity (C-TISTA (linear); circles), and DFT (cross marks) and lines represent PAPR$=1.5$dB (solid) and $5$dB (broken).} 

\label{fig_2}
\end{figure*}

\subsection{Complex-Valued Compressed Sensing}

We first consider complex-field CS related to spectrum sensing~\cite{spec} and angle-of-arrival detection~\cite{arrival}.

We consider an underdetermined linear system $\bm{y}=\bm{A}\bm{x}+\bm{w}$ of size $(n,m) = (240,120)$}. 
As sparse prior, we assume that each element of $\bm x$ follows the complex Gaussian-Bernoulli prior
$p(x)=(1-p)\delta(x) +p G(x;0,\sigma_\mathrm{x}^2)$ 
with $p=0.1$ and $\sigma_\mathrm{x}^2 = 1.0$.
This means that there are around 10\% non-zero elements in $\bm{x}$.
Each component of $\bm{A}$ follows ${\cal CN}(0, 1/{m})$ and
the noise variance of $\bm{w}$ is $\sigma^2 = 0.03^2$.

The soft thresholding function $\eta_{cs}(\cdot)$ is used in C-TISTA.
The number of mini-batches is $K=2000$ and the batch size is $L=200$ in its training process.

{As baselines, %of signal recovery performance, 
we perform AMP~\cite{AMP} and complex AMP (CAMP)~\cite{CAMP}.
Complex numbers are transformed to real numbers to execute AMP.
In addition, complex-field LISTA (C-LISTA) is also executed for comparison.
It is defined by
\begin{equation} 
\label{CLISTA_gradient}
\bm{s}^{(t + 1)} := \eta_{cs}(\bm B \bm{y} + \bm C \bm{s}^{(t)}; \tau_t),
\end{equation}
with trainable parameters $\bm{B}\in\mathbb{C}^{n\times m}$, $\bm{C}\in\mathbb{C}^{n\times n}$, and $\tau_t\in \mathbb{R}$ ($t=1,\dots,T$).
{For stable training and reasonable performance}, trainable parameters are {are shared with all iterations, i.e.,} $\bm B$ and $\bm C$ unlike the real-valued case~(\ref{eq_LISTA}).
}

Figure~\ref{fig_cs} shows normalized MSE (NMSE) performance as a function of the number of iterations $T$. %of algorithms.
It is found that the NMSE of C-TISTA decreases most rapidly.
C-TISTA reaches to NMSE$=$$-25$dB at $T=6$ and almost converges to $-26$dB at $T=8$.
AMP and CAMP show slower convergence reaching to around $-20$dB and 
$-21$dB at $T=15$, respectively.
Neglecting correlations of complex vectors by a complex-to-real transformation in AMP degrades 
signal recovery performance.
{Although C-LISTA shows better NMSE performance than AMP and CAMP,
C-TISTA outperforms C-LISTA in terms of convergence speed and convergent NMSE (about $-25$dB {by C-LISTA}) with faster training process.
This is because C-TISTA is more flexible thanks to trainable step sizes $\{\beta_t\}_{t=1}^T$ and a tunable DF-like function. 
It is emphasized that C-TISTA has only $4T$ trainable parameters 
while C-LISTA has $n^2+nm+T$ parameters in total.}
These results show that C-TISTA successfully recovers complex-valued sparse signals with fast convergence 
and improved accuracy.

\subsection{16-QAM Signal Detection in Clipped OFDM Systems}
We next examine a clipped OFDM system as a nonlinear system with discrete inputs.
{Amplitude clipping limits peaks of a time-domain signal containing the sum of contributions from multiple carriers.
It reduces the peak-to-average power ratio (PAPR) with low complexity~\cite{Han} and is useful to avoid signal saturation at an amplifier.}
However, it causes signal distortion that {degrades performance of the zero-forcing detector by} the discrete Fourier transformation (DFT).

The clipped OFDM system is formulated as the nonlinear system $\bm{y}\!=\!f_c(\bm{F}\bm{x})\!+\!\bm{w}$ of $n$ carriers.
We assume that each signal $x_i$ is uniformly chosen from 16-QAM signal points, i.e., $x_i\!=\!p\!+\!\im q$ ($p,q\!\in\!\{\pm 1,\pm 3\}$).
The matrix $\bm F$ is the $n\times n$ inverse DFT matrix 
{whose $(l,m)$-element is given by $\exp\left(\im {2\pi lm}/{n}\right)/{\sqrt{n}}$.}
%defined by
%\begin{equation}
%$f_{ki} = \exp\left(\im {2\pi ki}/{n}\right)/{\sqrt{n}} %\label{eq_ofdm}
%$ %\end{equation}
The function $f_c(z)$ is a clipping function defined by %which takes
%\begin{comment}
\begin{equation}
f_c(z) :=  
\begin{cases}
z & (|z|\le \alpha )\\
\alpha e^{\im \varphi(z)} & (|z|>  \alpha) %\nonumber
\end{cases}. \label{eq_clip}
\end{equation}
%\end{comment}
%$z$ ($|z|\le \alpha$) and $\alpha \exp(\im \varphi(z))$ otherwise,
%where $\alpha(>0)$ is called a clipping level.
%We assume that each component of $\bm{w}$ follows $\mathcal{CN}(w; 0,\sigma^2)$.
{The clipping level $\alpha(>0)$ is defined as the PAPR given by  $%\begin{equation}
10\log_{10} ({\max_{1\le k\le n} |\tilde{x}_k|^2}/{\mathsf{E}_{\bm{x}}\|\bm{Fx}\|_2^2 }) %\label{eq_papr}
$,} %\end{equation}
where $\bm{\tilde{x}}\! :=\! f_c(\bm{Fx})$.
Small PAPR indicates large distortion.
{Here, we set $n\!=\!128$ and PAPR$=\!1.5$dB or $5$dB as strongly clipped systems.}

As a shrinkage function of C-TISTA, we use 
the MMSE function (\ref{eq_sh}) based on the uniform distribution over the 16-QAM constellation.
In addition to C-TISTA, we also perform C-TISTA without nonlinearity to study influence of nonlinearlity.
C-TISTA without nonlinearity named C-TISTA (linear) has a gradient corresponding to $f_c(z)=z$. 
In the training process, we set $K=500$, $L=200$, and $T=10$.

Figure~\ref{fig_2} (a) shows the symbol error ratio (SER) performance as a function of the signal-to-noise ratio (SNR).
The results show that SERs of DFT and C-TISTA (linear) are largely affected by the value of PAPR.  
In contrast, C-TISTA shows nearly the same SER performance even when the clipping level is small.
It suggests that the signal distortion introduced by amplitude clipping is correctly compensated by C-TISTA.
In fact, compared with DFT and C-TISTA (linear), C-TISTA exhibits better detection performance: 
the gain of C-TISTA is about $2$dB when SER$=\!1.0\times 10^{-3}$ and PAPR$=$5dB, which becomes much larger when PAPR$=1.5$dB.

MSE performance is also important if 
a belief propagation decoder for LDPC codes is applied to the system
to deal with in-band distortion~\cite{Sori}.
Fig.~\ref{fig_2} (b) shows the SNR dependency of the MSE performance in the same setting.
The result shows that C-TISTA also detects transmitted signals with high accuracy. %in terms of the MSE.
These results suggest that C-TISTA can detect a discrete signal even in nonlinear inverse problems.

\section{Conclusion}\label{sec_sum}
In this paper, we propose C-TISTA for complex-field nonlinear inverse problems that have wide applications such as {complex-valued CS} and signal detection from nonlinear measurements. 
C-TISTA consists of a gradient step with Wirtinger derivatives {followed by a} shrinkage step with a trainable complex shrinkage function. {Its training is scalable and stable because it has a small number of trainable parameters.}
Numerical studies reveal that C-TISTA shows remarkable signal recovery performance in complex-valued CS and clipped OFDM systems.
These results indicate a promising potential of C-TISTA for wide range of inverse problems
in signal processing and wireless communications.

%\vfill\pagebreak

% References should be produced using the bibtex program from suitable
% BiBTeX files (here: strings, refs, manuals). The IEEEbib.bst bibliography
% style file from IEEE produces unsorted bibliography list.
% -------------------------------------------------------------------------
%\bibliographystyle{IEEEbib}
%\bibliography{strings,refs}

\appendix
\section{Gradient Descent by the Wirtinger Derivative}\label{sec_app}
In this Appendix, we briefly review Wirtinger derivative and 
derive a gradient descent method for solving the least mean square (LMS) problem for (\ref{eq_GLM}), i.e.,
\begin{equation}
\mathrm{Minimize}_{\bm{x}\in\mathbb{C}^n } g(\bm{x}) :=\frac{1}{2}\|\bm{y} - f(\bm{A}\bm{x})\|_2^2, \label{eq_LMS}
\end{equation}
without prior information.

A simple extension of the gradient descent method for a real-valued problem is not applicable 
to the complex-valued problem (\ref{eq_LMS}) because the function $g(\bm{x})$ does not have complex differentiability.
To overcome this difficulty, we here use Wirtinger derivative in complex analysis as an extension of partial derivative
with respect to real variables. 
For a function $f:\mathbb{C} \rightarrow \mathbb{C}$ and
${z}:={z}_{\mathrm{r}}+\im {z}_{\mathrm{i}}\in \mathbb C$ (${z}_{\mathrm{r}}=\mathrm{Re}(z), {z}_{\mathrm{i}}=\mathrm{Im}(z)$), we introduce a function $F:\mathbb{R}^2 \rightarrow \mathbb{C}$ such that 
$F({z}_{\mathrm{r}},{z}_{\mathrm{i}}) = f(z)$.
Wirtinger derivative is then defined  by
\begin{align}
\frac{\partial f(z)}{\partial z} &:= \frac{1}{2} \left[ \frac{\partial}{\partial {z}_{\mathrm{r}}}-\im \frac{\partial}{\partial {z}_{\mathrm{i}}} \right] F({z}_{\mathrm{r}},{z}_{\mathrm{i}}), 
\\
\frac{\partial f(z)}{\partial z^\ast} &:= \frac{1}{2} \left[ \frac{\partial}{\partial {z}_{\mathrm{r}}}+\im \frac{\partial}{\partial {z}_{\mathrm{i}}} \right] F({z}_{\mathrm{r}},{z}_{\mathrm{i}}), 
\end{align}
by using partial derivatives~\cite{Rem}.
For a complex vector $\bm{z}\in\mathbb{C}^n$, we define differential operators 
%with respect to $\bm{z}$ and $\bm{z}^\ast$ 
%\begin{align}
$\frac{\partial }{\partial \bm{z}} := \left( \frac{\partial}{\partial z_1},\dots, \frac{\partial}{\partial z_n}\right)^\mathrm{T}$, and 
$\frac{\partial }{\partial \bm z^\ast} := \left( \frac{\partial}{\partial z_1^\ast},\dots, \frac{\partial}{\partial z_n^\ast}\right)^\mathrm{T}$.
%\end{align}
It is known that the steepest descent direction 
 is given by $-\partial f/ \partial \bm{z}^\ast$~\cite{LMS}, not $-\partial f/ \partial \bm{z}$.
We thus calculate $\nabla g(\bm{x}):=-\partial g/ \partial \bm{x}^\ast$ of  (\ref{eq_LMS}) to solve the LMS problem
by gradient descent.

We first calculate the Wirtinger derivative with respect to the first variable $x_1^\ast$
based on the assumption that the function $f$ is applied to a vector component-wisely. We have  
\begin{align}
\frac{\partial}{\partial x_1^\ast}g(\bm{x}) 
=& -\frac{1}{2}\sum_{i=1}^m \left[ \{y_i - f(\bm{A}_{i,:}\bm{x})\}^\ast \frac{\partial}{\partial x_1^\ast} \left(f(\bm{A}_{i,:}\bm{x})\right) \right.\nonumber\\
&\left.\qquad\qquad+\{y_i - f(\bm{a}_{i,:}\bm{x})\} \frac{\partial}{\partial x_1^\ast} \left(f^\ast(\bm{A}_{i,:}\bm{x})\right) \right] \nonumber\\
=& -\frac{1}{2}\sum_{i=1}^m A_{i1}^\ast \left[ \{y_i - f(\bm{A}_{i,:}\bm{x})\}^\ast \frac{\partial f}{\partial z^\ast}(\bm{A}_{i,:}\bm{x})\right.\nonumber\\
&\left.\qquad\qquad\quad+\{y_i - f(\bm{A}_{i,:}\bm{x})\} \frac{\partial f^\ast}{\partial z^\ast}(\bm{A}_{i,:}\bm{x})\right],
\label{eq_wir0}
\end{align}
where $\bm{A}_{i,:}:=(A_{i1},\dots, A_{in})$ ($1\le i\le m$) of the matrix $\bm A$ and 
$\frac{\partial f}{\partial z^\ast}\left(\bm{u}\right):= \left.\frac{\partial f(z)}{\partial z^\ast}\right|_{z=\bm{u}}$.
We use a chain rule~\cite{Rem}{, i.e.,
\begin{equation}
\frac{\partial}{\partial z^\ast}(f_1\circ f_2) = \left( \frac{\partial f_1}{\partial z}\circ f_2 \right) \frac{\partial f_2}{\partial z^\ast}
+\left( \frac{\partial f_1}{\partial z^\ast}\circ f_2 \right) \frac{\partial f_2^\ast}{\partial z^\ast},
\end{equation}
for functions $f_1,f_2:\mathbb{C}\rightarrow\mathbb{C}$,}
 and an identity $\partial z/ \partial z^\ast=\partial z^\ast/ \partial z=0$ in the second line of (\ref{eq_wir0}).

Combining it with derivatives with respect to other variables, we finally obtain
\begin{align}
\nabla g(\bm{x}) 
&=-\frac{1}{2}\bm{A}^\lH
\left[ \{\bm{y} - f(\bm{A}\bm{x})\}^\ast \odot \frac{\partial f}{\partial z^\ast}(\bm{A}\bm{x})\right.\nonumber\\
&\left.\qquad\qquad\quad+\{\bm{y} - f(\bm{A}\bm{x})\} \odot \frac{\partial f^\ast}{\partial z^\ast}(\bm{A}\bm{x})\right],  \label{eq_nabla}
\end{align}
where 
$\frac{\partial f}{\partial z^\ast}\left(\bm{u}\right):= \left.\frac{\partial f(z)}{\partial z^\ast}\right|_{z=\bm{u}}$.
It corresponds to (\ref{eq_tista_grad}) in the gradient step of C-TISTA when $\bm V=\bm A^\lH$.

Based on the gradient~(\ref{eq_nabla}), the update rule of the gradient descent method is given by 
\begin{equation}
\bm{x}^{(t+1)} = \bm{x}^{(t)}-2\beta \nabla g(\bm{x}^{(t)}) 
%{\beta} \bm{A}^\lH \left[ \{\bm{y} - f(\bm{A}\bm{x}^{(t)})\}^\ast \odot \frac{\partial f}{\partial z^\ast}(\bm{A}\bm{x}%^{(t)})
%+\{\bm{y} - f(\bm{A}\bm{x}^{(t)})\} \odot \frac{\partial f^\ast}{\partial z^\ast}(\bm{A}\bm{x}^{(t)})\right],  
\label{eq_gd_glm}
\end{equation}
with a step-size parameter $\beta(>0)$ and an initial value $\bm{x}^{(1)}$.
The factor $2$ in (\ref{eq_gd_glm}) is necessary to keep consistency with the real-valued case. 

It would be useful to simplify~(\ref{eq_nabla}) for several special cases.
{For} a linear model in which $f(z)=z$, we have
\begin{equation}
\nabla g(\bm{x}) =-\frac{1}{2}\bm{A}^\lH (\bm{y} - \bm{A}\bm{x}).  \label{eq_nabla_lin}
\end{equation}
This corresponds to the gradient descent step in ISTA (\ref{ISTA_gradient})
for real-valued linear systems.

If $f(z)$ is an analytic function, we have
\begin{equation}
\nabla g(\bm{x}) =-\frac{1}{2}\bm{A}^\lH \left[\{\bm{y} - f(\bm{A}\bm{x})\}\odot f'(\bm{A}\bm{x})\right],  \label{eq_nabla_2}
\end{equation}
where $f'(z)$ is a complex derivative with respect to $z$. Similarly, for a real-valued system 
in which $\bm{A}\in\mathbb{R}^{m\times n}$, $\bm{x}, \bm{w}\in\mathbb{R}^{n}$, and $f:\mathbb{R}\rightarrow \mathbb{R}$, 
we obtain
\begin{equation}
\nabla g(\bm{x}) =-\frac{1}{2}\bm{A}^T \left[\{\bm{y} - f(\bm{A}\bm{x})\}\odot f'(\bm{A}\bm{x})\right]. \label{eq_nabla_3}
\end{equation}

\end{document}